\documentclass[a4paper,fleqn,usenatbib]{mnras}
\usepackage{mathptmx}
 \usepackage{txfonts}
 \usepackage{textcomp}

\usepackage[T1]{fontenc}
\usepackage{ae,aecompl}

\usepackage{graphicx}	
\usepackage{multicol}        
\usepackage{pdflscape}	
\usepackage{multirow}

\title[The composition of Mars Trojan asteroids]{The olivine-dominated composition of the Eureka family of Mars Trojan asteroids}

\author[G. Borisov et al.]{
G. Borisov,$^{1,2}$\thanks{E-mail: gbb@arm.ac.uk (GB)}
A. Christou,$^{1}$\thanks{E-mail: aac@arm.ac.uk (AC)}
S. Bagnulo,$^{1}$
A. Cellino,$^{3}$
T. Kwiatkowski$^{4}$,
A. Dell'Oro$^{5}$
\\
$^{1}$Armagh Observatory and Planetarium, College Hill, Armagh BT61 9DG, UK\\
$^{2}$Institute of Astronomy and National Astronomical Observatory, 72, Tsarigradsko Chauss\'ee Blvd., BG-1784 Sofia, Bulgaria\\
$^{3}$INAF - Osservatorio Astrofisico di Torino, via Osservatorio 20, I-10025 Pino Torinese (TO), Italy\\
$^{4}$Astronomical Observatory, Adam Mickiewicz University, ul. S\l{}oneczna 36, PL-60-286 Pozna\'{n}, Poland\\
$^{5}$INAF - Osservatorio Astrofisico di Arcetri, Largo E. Fermi 5, I-50125, Firenze, Italy\\
}

\date{Accepted 2016 November 23. Received 206 November 22; in original form 2016 October 18}

\pubyear{2017}

\begin{document}
\label{firstpage}
\pagerange{\pageref{firstpage}--\pageref{lastpage}}
\maketitle

\begin{abstract}
We have used the XSHOOTER echelle spectrograph on the European South Observatory (ESO) Very Large Telescope (VLT) 
to obtain UVB-VIS-NIR (ultraviolet-blue (UVB), visible (VIS) and near-infrared (NIR)) 
reflectance spectra of two members of the Eureka family of L5 Mars Trojans in order to test a genetic relationship to 
Eureka. In addition to obtaining spectra, we also carried out {\it VRI} photometry of one of the VLT targets using 
the 2-m 
telescope at the Bulgarian National 
Astronomical Observatory -- Rozhen and the two-channel focal reducer. 
We found 
that these asteroids belong to the olivine-dominated A, or S$_{\rm a}$, taxonomic class. As Eureka itself is also 
an olivine-dominated asteroid, it is likely that {\it all} family asteroids share a common origin and composition. 
We discuss the significance of these results in terms of the origin of the martian Trojan population.
\end{abstract}

\begin{keywords}
planets and satellites: individual: Mars -- minor planets, asteroids: individual: Trojan asteroids -- techniques: imaging spectroscopy -- techniques: photometric
\end{keywords}



\section{Introduction}

The so-called Mars Trojans are asteroids located in the L4 or L5 Lagrangian points of Mars. 
They are thought to have been there since the very early phases of the Solar System's history. 
There were nine confirmed Martian Trojans in total as of 2015, of which eight were at L5 and 1 at L4.
Occupants of the Mars Trojan clouds might represent a small set of survivors of an early generation 
of planetesimals from which the inner Solar System was built.
The long-term stability of the Mars Trojan clouds has been extensively investigated by 
\citet{Scholl.et.al2005}. According to these authors, in the L4 and L5 regions of Mars, there are combinations 
of orbital eccentricity and inclination that are stable over time-scales comparable to the Solar System's age. 
Of the eight L5 Trojans, seven (including Eureka) form the Eureka family, whose existence has been pointed out 
by \citet{Christou2013} and \citet{deLaFuenteMarcoses2013}. Due to its compactness, 
this is probably a genetic family rather than a random grouping of orbits. 
It formed sometime in the last Gyr \citep{Cuk.et.al2015}. The members of the Eureka asteroid family are also stable Trojans.
For this reason, the discovery that asteroid (5261) Eureka belongs 
to the rare A taxonomic class \citep{Rivkin.et.al2007, Lim.et.al2011} is very interesting. The spectral reflectance 
properties of A-class asteroids have been interpreted as diagnostic of an overall composition rich in the 
mineral olivine [$(\mbox{Mg}^{++},\mbox{ Al}^{++})_2\mbox{SiO}_4$], a magnesium iron silicate that is a primary 
component  of the Earth's mantle and -- supposedly -- of the other terrestrial planets. The low abundance of olivine-rich 
objects among asteroids is an old conundrum. In particular, the existence of metal meteorites suggests that during 
the early phases of the Solar System's history, several planetesimals reached sizes sufficient for complete thermal 
differentiation due to the heat produced by the decay of short-lived isotopes, primarily \mbox{${\rm ^{26}Al}$}. 
Metal meteorites are interpreted in this scenario as fragments of the metal-rich cores of such differentiated bodies, 
exposed after their complete collisional disruption. In this context, one should expect that olivine-rich meteorites 
and asteroids should be common because olivine is a primary component of the mantles of differentiated bodies. 
However, this prediction is not borne out of the observations. Asteroids belonging to the olivine-rich A class, or 
even to the fairly similar S$_{\rm a}$ sub-class of the S taxonomic complex \citep{BusBinzel,DeMeoDB}, are quite 
rare. At the end of the 1990s, several authors proposed that the original olivine-rich asteroids were 'battered to bits' 
by collisions and had thus disappeared long ago \citep{Burbine.et.al1996,Chapman1997}. Whatever the cause of 
such disappearance, collisional evolution or, perhaps more likely, massive removal of the original planetesimals 
accreted in the main belt, as predicted by the Grand Tack and Nice models of early evolution of the Solar System 
\citep{Walsh.et.al2011,Walsh.et.al2012}, only a small number of survivors belonging to the A class can be found 
among the current asteroid population. As a consequence, the discovery of an A-class asteroid confined in a special 
dynamical environment within the terrestrial planet region suggests that the other members of the Eureka family 
also deserve a careful investigation. Based on these considerations, we began an observing campaign to obtain 
information on the spectral reflectance properties of these objects. This is a challenging task because, in spite of their 
relative proximity to the Earth, Mars Trojans are faint. Their observation therefore requires the use of 
large-aperture instruments.
 
In our investigation, we used XSHOOTER, the first European South Observatory (ESO) second-generation instrument developed for the Very Large 
Telescope \citep[VLT;][]{XSHOOTER}, which is currently mounted on VLT Unit 2 Kueyen. XSHOOTER is a spectrograph capable of 
obtaining spectra from 300 to 2480\,nm in a single shot at high spectral resolution (from 3000 up to 10\,000). In the 
following sections we present the spectra we obtained for two members of the Eureka family, and compare them with 
the available reflectance spectra for Eureka itself.

We also obtained some limited spectrophotometric information by carrying out multiband {\it VRI} photometry. 
We used the two-channel focal reducer -- Rozhen (FoReRo2) 
of the 2m \,Ritchey-Chr\'{e}tien-Coud\'{e} telescope at the Bulgarian National Astronomical Observatory (BNAO) 
to obtain $(V-R)$ and $(V-I)$ colours of a member of the Eureka family, also one of our two targets observed with
XSHOOTER. The colour indices were compared with the results obtained by \citet{Taxonomy}, and with the 
average values of the colour indices for different asteroid taxonomic classes.

This paper is organised as follows. In Section~\ref{obs}, we present our VLT and Rozhen observations. The adopted 
procedures for data reduction are described in Section~\ref{datared}, and the results in Section~\ref{results}, separately 
for spectroscopy and multiband photometry. Our main conclusions and a general discussion of our results is given in Section~\ref{blabla}.

\section{Observations}
\label{obs}
We used XSHOOTER during two nights, 
FoReRo2 during two nights as well and ACAM for 1 night. The objects that were observed are as follows: 
XSHOOTER: (385250) 2001 DH47, (311999) 2007 NS2 and spectral solar analogue star HD 67010; 
FoReRo2: (385250) 2001 DH47, (289) Nenetta 
[an asteroid with olivine-dominated surface \citep{Sanchez}]
and Stetson standard field L101;
ACAM: (3819) Robinson 
[an asteroid with olivine-dominated surface \citep{Sanchez}]
and Stetson standard field L104.
Details of our observations are presented in Tables~\ref{obsxs}--\ref{obswht}.

\subsection{Spectroscopy}
\subsubsection{XSHOOTER}
This instrument is an echelle spectrograph with an almost fixed spectral setup. The observer can choose 
between SLIT (and slit width) and IFU (integral field unit) modes. Here we used the SLIT mode. A detailed 
description of the instrument is available at \citet{XSHOOTER} and ESO's webpage\footnote{http://www.eso.org/sci/facilities/paranal/instruments/xshooter.html}. 
This spectrograph has the ability to simultaneously obtain data over the entire 300--2480\,nm spectral 
range by splitting the incoming light from the telescope into three beams, each sent to a different arm: 
ultraviolet--blue (UVB), visible (VIS), and near-infrared (NIR). Using two dichroic filters, the light is first 
sent to the UVB arm, then to the VIS arm, and finally the remaining light arrives at the NIR arm. The 
disadvantage of this choice of optical light path is the high thermal background in the K-region of the 
NIR spectrum. The observations are presented in Table~\ref{obsxs}. Observations were done 
in nodding mode to facilitate subsequent sky signal estimation and subtraction.

\begin{table*}
\makebox[0pt][c]{\parbox{0.8\textwidth}{%
\begin{minipage}[b]{\linewidth}\centering 
\caption[Observations - XSHOOTER@VLT]{Observations - XSHOOTER@VLT.}
\label{obsxs}
\begin{tabular}{lcccccr}
\hline 
\hline
\multirow{2}{*}{Object} & \multirow{2}{*}{Date} & \multirow{2}{*}{UT} &  \multirow{2}{*}{Airmass}   & Apparenet    & Spectral  & Exposure time\\
                                    &                                   &                                &                                            & magnitude   &  arm        & (s)\\
\hline
                                   &                      & 03:04 & 1.174 &           & UVB & 3040 \\
(385250) 2001 DH47& 2016 February 02 &\textbar&\textbar& 19.00 & VIS & 3280 \\
                                   &                      & 04:07 & 1.377 &           & NIR & 3600 \\
\hline
                 &                      &           &           &           & UVB & 40 \\
HD 67010 & 2016 February 02 & 05:15 & 1.158 & 8.69   & VIS & 30 \\
                 &                      &           &           &           & NIR & 100 \\
\hline
                                &                       & 01:37 & 1.022 &           & UVB & 3040 \\
(311999) 2007 NS2& 2016 March 02 &\textbar&\textbar& 19.05 & VIS & 3280 \\
                                &                       & 02:21 & 1.367 &           & NIR & 3600 \\
\hline
                 &                      &           &           &           & UVB & 40 \\
HD 67010 & 2016 March 02 & 03:30 & 1.168 & 8.69   & VIS & 30 \\
                 &                      &           &           &           & NIR & 100 \\
\hline
\end{tabular}
\end{minipage}
}}
\end{table*}

\subsection{\textbf{\textit{VRI}} Photometry}
\subsubsection{FoReRo2} 
The instrument is a two-channel focal reducer that adapts the imaging elements of the detector to 
the characteristic size of the object or the seeing disk. FoReRo2 was built mainly for observations 
of cometary plasma but has been proved suitable for many other tasks \citep{FoReRo2}. Behind the RC 
(Cassegrain) focus, the light beam is recollimated by a lens collimator. A dichroic mirror reflects the 
blue part of the spectrum and transmits the red part. For each channel, camera lenses form reduced 
images of the Cassegrain focal plane, which are recorded by two CCD systems. Filters are placed 
into the parallel beam after colour separation. For this investigation we used {\it VRI} standard filters; 
therefore, only the red channel of the instrument was in operation. The observations are 
presented in Table~\ref{obs2m}.

\begin{table*}
\makebox[0pt][c]{\parbox{0.8\textwidth}{%
\begin{minipage}[b]{\linewidth}
\centering 
\caption[Observations - FoReRo2@2mRCC]{Observations - FoReRo2@2mRCC.}
\label{obs2m}
\begin{tabular}{lcccccr}
\hline 
\hline
\multirow{2}{*}{Object} & \multirow{2}{*}{Date} & \multirow{2}{*}{UT} &  \multirow{2}{*}{Airmass}   & Apparenet    & \multirow{2}{*}{Filter} & Exposure time\\
                                    &                                   &                                &                                            & magnitude   &             & (s)\\
\hline

                                   & 2016 February 06 & 22:00 & 1.30 &               & \it I & $10\times300$ \\
(385250) 2001 DH47&     \textbar     &\textbar&\textbar& 18.55 & \it R & $10\times300$ \\
                                   & 2016 February 07 & 01:00 & 1.50 &               & \it V & $10\times300$ \\
\hline
                       &                      & 20:00 & 1.15 &               & \it I & $3\times300$ \\
(289) Nenetta & 2016 February 06 &\textbar&\textbar& 14.13 & \it R & $3\times300$ \\
                       &                      & 21:00 & 1.25 &               & \it V & $3\times300$ \\
\hline
                      &                      & 19:50 & 2.10 & 12.64     &  \it V & $3\times(5\times60)$ \\
Stetson L101 & 2016 February 06 & 21:30 & 1.50 & \textbar &  \it R & $3\times(5\times60)$ \\
                     &                        & 23:00 & 1.30 & 20.64    &  \it I & $3\times(\times60)$ \\
\hline
\end{tabular}
\end{minipage}
}}
\end{table*}

\subsubsection{ACAM} 
The ACAM instrument of the 4.2-m William Herschel Telescope (WHT) at the Observatorio 
del Roque de los Muchachos (La Palma, Spain) was used to obtain SDSS photometry of 
asteroid (3819) Robinson. The observations are presented in Table~\ref{obswht}.

\begin{table*}
\makebox[0pt][c]{\parbox{0.8\textwidth}{%
\begin{minipage}[b]{\linewidth}
\centering 
\caption[Observations - ACAM@WHT]{Observations - ACAM@WHT.}
\label{obswht}
\begin{tabular}{rcccccr}
\hline 
\hline
\multirow{2}{*}{Object} & \multirow{2}{*}{Date} & \multirow{2}{*}{UT} &  \multirow{2}{*}{Airmass}   & Apparenet    & \multirow{2}{*}{Filter} & Exposure time\\
                                    &                                   &                                &                                            & magnitude   &             & (s)\\
\hline
                            &                      & 06:10 & 1.94 &               & \it g\textquotesingle & $3\times80$ \\
(3819) Robinson & 2016 March 11 &\textbar& 2.02& 16.36     & \it r\textquotesingle & $3\times60$ \\
                           &                       & 06:25 & 2.12 &               & \it i\textquotesingle & $3\times60$ \\
\hline
                      &                      & 02:30 & 1.145 & 12.75     &  \it g\textquotesingle & $(4\times5)$ and $(3\times15)$ \\
Stetson L104 & 2016 March 11 &           &           & \textbar  &  \it r\textquotesingle & $(4\times5)$ and $(3\times10)$ \\
                     &                        & 06:00 & 1.853 & 21.12    &  \it i\textquotesingle & $(4\times5)$ and $(3\times15)$ \\
\hline
\end{tabular}
\end{minipage}
}}
\end{table*}

\section{Data reduction}
\label{datared}

\subsection{XSHOOTER}
To reduce the XSHOOTER spectra we first processed the data through the ESO 
\textsc{esoreflex} pipeline version 2.6.8. All spectra were reduced under the assumption of  point-like sources and 
had their instrument signature removed, i.e. de-biased, flat-fielded,  wavelength-calibrated, order-merged, 
extracted, sky-subtracted and, finally, flux-calibrated. 

To achieve a signal-to-noise ratio (S/N) sufficient for scientific analysis the reduced 1D spectra for both the asteroid 
and solar analogue standard stars were then rebinned in 20\,nm steps. Subsequently, the spectrum of the 
asteroid was divided by the solar analogue spectrum, the result normalised to unity at 550\,nm and 
again smoothed using a floating average with different wavelength steps in the VIS and NIR regions.

Finally, we carried out running sigma clipping of the data with a 20\,nm  window in wavelength and a 
3\,$\sigma$ criterion. During this stage we excluded sections of the spectra affected by high telluric line 
contamination in the following wavelength intervals: 0.90--1.00, 1.35--1.50 and 1.80--1.90\,${\rm \mu}$m, as 
suggested by \citet{teluric1} and \citet{teluric2}.







\subsection{FoReRo2}
All imaging data had their instrument signature removed as well by de-biasing and flat-fielding. The images 
in I were de-fringed using the median I image combined from all I images for the night. Standard \textsc{daophot} 
aperture photometry with aperture size 2$\times$full width at half-maximum (FWHM) was performed. Aperture correction, which was 
measured through a large aperture using the growth-curve method \citep{ApCor}, was applied in order to 
place the instrumental magnitudes of the observed point source objects, which were measured through a 
small aperture (2$\times$FWHM), on the same system as those of the standard stars. Stetson standard field 
L101, observed on three different airmasses, was used for obtaining extinction coefficients in each filter. Linear 
regression fits for magnitude--magnitude and colour--colour calibration were performed as well. Then, the instrumental 
magnitudes of the asteroid targets were absolutely calibrated using coefficients from those fits, and, finally, 
their true $(V-R)$ and $(V-I)$ colours were computed.

\subsection{ACAM} 
All imaging data had their instrument signature removed as well by de-biasing and flat-fielding. The same 
calibration procedure as the one for FoReRo2 was performed using Stetson standard field L104. The 
photometry performed in SDSS-{\it g\textquotesingle r\textquotesingle i\textquotesingle}
filters was converted to {\it VRI} using relations presented by \citet{SDSS-UBVRI}.

\section{Results}
\label{results}

\subsection{Spectroscopy}

\begin{figure}
\includegraphics[angle=90,width=\columnwidth]{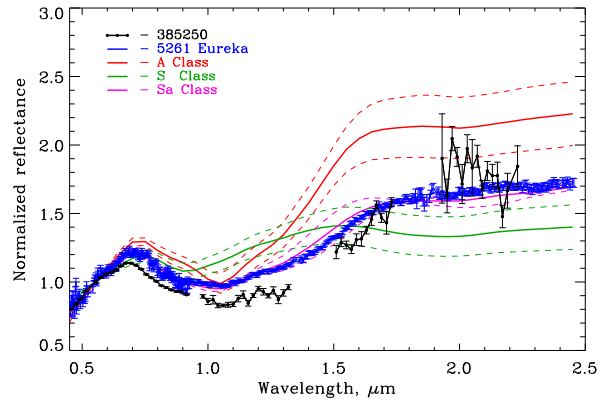}
\caption{(Colour online) Reflectance spectrum of the asteroid (385250) 2001 DH47 (black dotted) compared with average spectra for 
A (red), S (green) and S$_{\rm a}$ (purple) taxonomic classes. The spectrum of (5261) Eureka is overplotted in blue. 
}
\label{385250RefSpCl}
\end{figure}

\begin{figure}
\includegraphics[angle=90,width=\columnwidth]{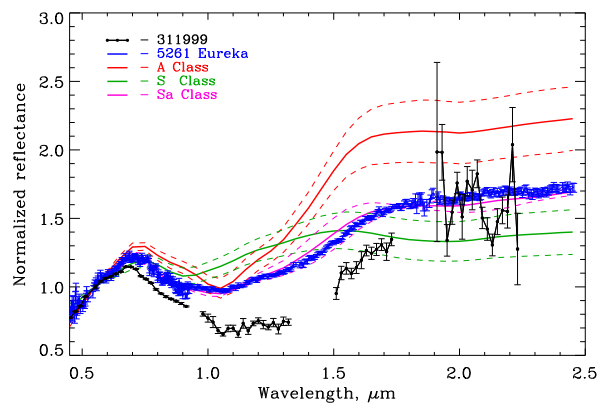}
\caption{(Colour online) Reflectance spectrum of the asteroid (311999) 2007 NS2  (black dotted line) compared with average spectra for 
A (red), S (green) and S$_{\rm a}$ (purple) taxonomic classes. The spectrum of (5261) Eureka is overplotted in blue. 
}
\label{311999RefSpCl}
\end{figure}

\begin{figure}
\includegraphics[angle=90,width=\columnwidth]{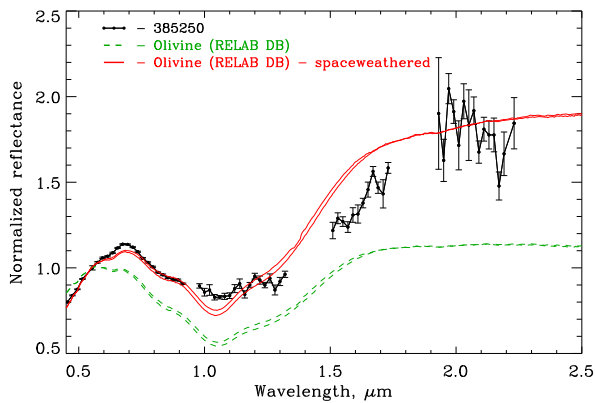}
\caption{Reflectance spectrum of asteroid (385250) 2001 DH47 (black line with points) compared with two olivine spectra 
from \textsc{relab} (green dashed line) and the same spectra divided by a slope to simulate space-weathering (red line). 
}
\label{385250RefSpLab}
\end{figure}

\begin{figure}
\includegraphics[angle=90,width=\columnwidth]{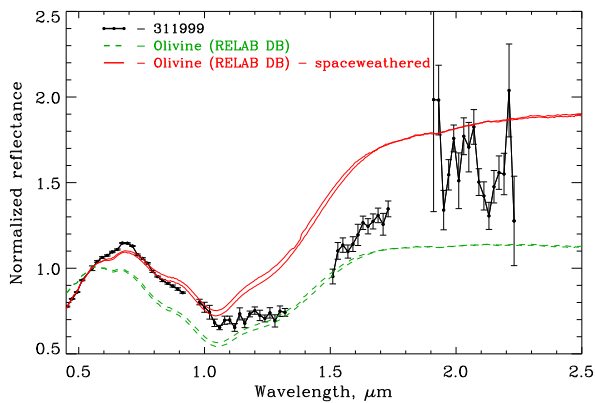}
\caption{Reflectance spectrum of the asteroid (311999) 2007 NS2 (black line with points) compared with two olivine spectra 
from \textsc{relab} (green dashed line) and the same spectra divided by a slope to simulate space-weathering (red line). 
}
\label{311999RefSpLab}
\end{figure}

The resulting XSHOOTER spectra are represented by black lines in 
Figs~\ref{385250RefSpCl}--\ref{311999RefSpLab}.

On Figs~\ref{385250RefSpCl} and~\ref{311999RefSpCl}, we show, apart from the reflectance spectra 
of our two targets, a spectrum  of (5261) Eureka (blue) obtained on 2015May 19 
\cite[][spectrum available online at {\tt http://smass.mit.edu/minus.html - sp41}]{SpeX} and average spectra for 
the A (red), S (green) and S$_{\rm a}$ (purple) classes in the \citet{DeMeoDB} taxonomy produced from data 
available in the PDS data base \cite[EAR-A-VARGBDET-5-BUSDEMEOTAX-V1.0;][]{SpDB} as point-by-point 
means of 6, 144 and 2 asteroids respectively. We find a 1\,$\mu$m absorption feature for 
both Eureka family members where the reflectance minimum is located slightly longward of 1\,$\mu$m. 
This is also the case for the A- and S$_{\rm a}$-type spectra and a diagnostic feature of an olivine-rich surface. 

The regions with high telluric lines contamination, excluded from our analysis, (see Section~\ref{datared}), do not affect  
distinguishing between the S, A and S$_{\rm a}$ classes, as the 1\,$\mu$m feature in S-class spectra is below 
0.9\,$\mu$m. We also note  the higher and flatter reflectance of the S-class spectrum in the region of 1.0-1.5\,$\mu$m, which distinguishes it 
from the other spectra.

Overall, the two asteroid spectra are in better agreement with an S$_{\rm a}$ than an A taxonomy, both in the visible and 
the infrared (IR). Eureka, also an S$_{\rm a}$-type asteroid in the \citeauthor{DeMeoDB} classification scheme\footnote{The observed 
coincidence between the S$_{\rm a}$ spectrum and that of Eureka is not accidental; this asteroid is one of only two objects 
whose spectra define this taxonomic class}, appears to be somewhat more reflective than 311999 and 385250 
from $\sim$0.6 up to at least 1.3\,$\mu$m. 
Eureka's 1\,$\mu$m feature is unique among A/S$_{\rm a}$-types. It is the only object with that particular band shape. 
The spectra of the two new objects, which we classify as S$_{\rm a}$, though they have a lower S/N ratio, appear consistent with Eureka's.
At 2$\mu$m the reflectivities of the three objects are indistinguishable 
from each other; although possibly a consequence of the lower S/N ratio, we believe it to be a real feature of 
the spectrum.

Seeking to strengthen the case for the presence of olivine, we compared our spectra with a number of laboratory olivine 
spectra from the \textsc{relab} data base. Exposure of asteroid surfaces to the space environment changes their optical properties. 
Therefore we applied a simple space weathering correction to the laboratory spectra prior to the 
comparison by dividing the olivine \textsc{relab} spectra with a first order polynomial. 
Through visual inspection, we found that spectra MS-CMP-042-A and MS-CMP-014 (green lines in Figs~\ref{385250RefSpLab} and~\ref{311999RefSpLab}
altered by our space-weathering correction, red lines) 
are reasonable matches to the asteroid spectra. The first laboratory spectrum corresponds to unprocessed 
pure olivine; the second, also of pure olivine, is altered by laser irradiation to simulate micrometeoroid impacts. 
Although it is difficult to obtain a good match over the entire wavelength interval covered by our spectra, 
by restricting ourselves to the visible part. 
and the simple space weathering model, we find that our best match (red lines) fits the visible 
part of our spectra quite well, particularly the minor absorption features around 0.63 and 0.8\,$\mu$m.
Progressing further along this line of investigation probably requires a more refined model of space-weathering effects, 
which is beyond the scope of this paper.

\subsection{\textbf{\textit{VRI}} Photometry}
Visible colour photometry can be used as a consistency check and also to eliminate candidate spectroscopic 
classes. Its advantage is that it is generally applicable to fainter objects than spectroscopy. On the other hand, 
unambiguous taxonomic classification and mineralogical interpretation typically requires detailed knowledge of 
the reflectance as a function of wavelength as well as extending observations into the IR. Our purpose here is 
to identify, through direct measurements, the domain that Eureka, its family members and asteroids of a similar 
mineralogical composition occupy in colour space with a view to possible future observations of the fainter 
Eureka family asteroids.

The $(V-R)$ and $(V-I)$ colour indices for one of the objects, (385250) 2001 DH47, were derived from three series 
of 10 exposures, acquired in {\it I}, {\it R}, and {\it V} filters with FoReRo2 at BNAO.  As some images were of inferior 
quality, only eight exposures in each filter were used, from which average values were computed.

Because of the faintness of the asteroid target, we used relatively long exposures of 300\,s. This resulted in a duty cycle 
of 318\,s and extended the observing run to 3\,h. On 2016 January 8 we carried out photometric observations of (385250) 2001 DH47. A rotation period of $P\approx 4.0 \pm 0.8$\,h and a peak-to-peak amplitude of 0.6\,mag were derived \citep{rot}. The asteroid light variation could potentially introduce systematic effects in the colour indices.

We used those parameters to understand the effect of the asteroid lightcurves in determining the colour indices obtained 
from the Rozhen data.  Due to the relatively high uncertainty in the rotation period $P$ and a long time interval 
between the January 8 and February 6 observations, we could not determine accurate rotation phases for our exposures. 
In addition, the January 8 observations were performed at the solar phase angle of 30.$\!^\circ$4, while on February 6 the phase angle 
was 8.$\!^\circ$2.  Assuming the classic amplitude--phase relationship with $m=0.02$\,mag deg$^{-1}$ \citep{amplitude}, 
we estimated the lightcurve amplitude of (385250) 2001 DH47 on February 6 to be 0.4\,mag.

\begin{figure}
\includegraphics[angle=90,width=\columnwidth]{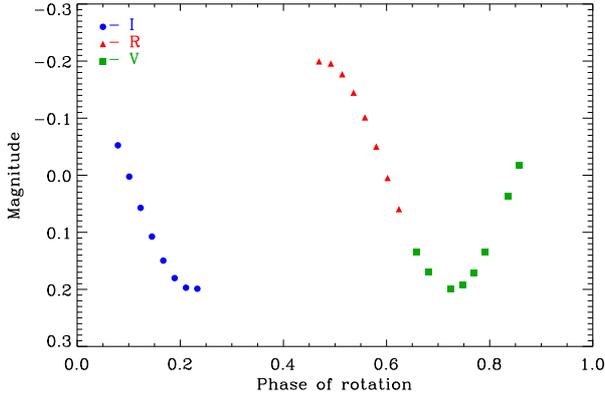}
\caption{A simulated lightcurve of (385250) 2001 DH47 with an assumed sinusoidal brightness variation, rotation period of 4\,h 
and a peak-to-peak amplitude of 0.4\,mag. While the rotation phase of the first point is arbitrary, data intervals reflect actual times 
of the exposures in the {\it I}, {\it R}, and {\it V} filters.}
\label{a2}
\end{figure}

A simulated lightcurve of (385250) 2001 DH47, with a simple sinusoidal shape, rotation period $P=4$\,h, and peak-to-peak 
amplitude $A=0.4$\,mag, is presented in Fig.~\ref{a2}, with the relative times of the {\it I}, {\it R}, and {\it V} exposures superimposed on the lightcurve. 
The rotation phase of the first point is arbitrary, but the intervals between consecutive points reflect the actual timings 
of the exposures.  In this model, the average brightness in the {\it I}, {\it R}, and {\it V} filters would be $0.084, -0.080$, and $0.102$, 
respectively, and the colour indices would be \mbox{$(V-I)=0.018$\,mag}, \mbox{$(V-R)=0.183$\,mag}.  In reality, the rotation phase of the 
first point of the observing sequence is unknown, so we repeated the computations with phase shifts from 0.02 to 1 in 
steps of 0.025.  Also, for each phase shift, we computed colour indices while changing the rotation period from 3.2\,h 
to 4.8\,h, in steps of 0.1\,h. In this way, we obtain $39\times 17= 663$ pairs of colour indices $(V-I)$, and $(V-R)$, 
which allowed us to estimate the systematic uncertainty of the colour indices computed from the Rozhen observations due to asteroid brightness variation. 

To estimate the statistical uncertainties resulting from the random scatter of the measurements we used formal 
sigma values given by the \textsc{daophot} package for each single $V$, $R$, $I$ instrumental magnitude. These were then 
propagated to final estimates of the $(V-I)$ and $(V-R)$ colours, resulting in sigma values of $0.05$ and $0.04$\,mag, respectively.  The former 
is slightly greater than the latter due to the greater noise of the {\it I}-filter measurements.  

\begin{figure}
\includegraphics[width=\columnwidth]{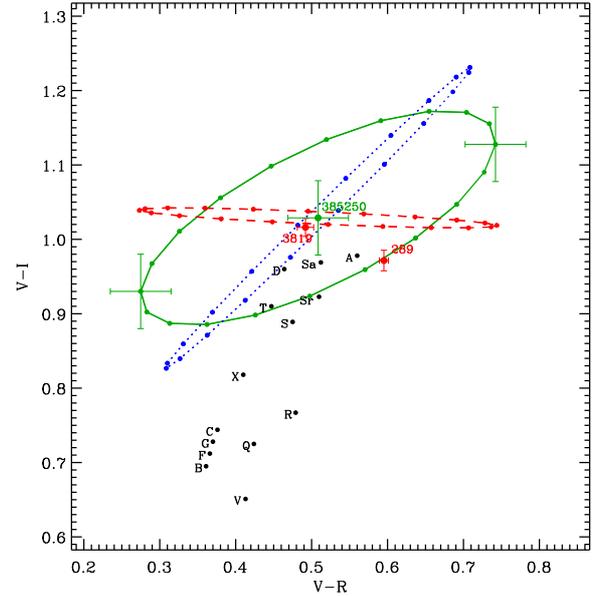}
\caption{$(V-R)$/$(V-I)$ colour diagram of the Eureka family member (385250) 2001 DH47 as well as A-class asteroids (289) Nenetta 
        and (3819) Robinson, versus mean colours of the different taxonomic classes discussed in the 
        text.
        The three ellipses shown in the plot represent the systematic uncertainty 
        in the colours of (385250) 2001 DH47 due to the lightcurve effect. These are shown 
        for the rotation period solution ($P=4$\,h; green line), as well as for two extreme cases: 
        $P=3.2$\,h (blue dotted line) and $P=4.8$\,h (red dashed line). If our measurements were free from the random error, our estimated 
        colours for this asteroid should be located on one of the eclipses. 
        Note that the random uncertainties (shown by error bars) are smaller than this 
        systematic effect.}
\label{VRI}
\end{figure}

Fig.~\ref{VRI} presents the $(V-R)$ and $(V-I)$ colour indices of (385250) 2001 DH47, compared with the average colours of all taxonomic 
classes in the \citet{TholenPhD} classification scheme as given in \citet{Colours} as well as the S$_{\rm a}$ and S$_{\rm r}$ 
classes in the \citet{BusBinzel} classification.  The systematic uncertainty due to the lightcurve is shown as an ellipse. The 
random measurement uncertainties of the colour indices are represented by the error bars.

S$_{\rm a}$ and S$_{\rm r}$ classes in the \citet{BusBinzel} taxonomy -- distinct from the classes of the same name in the \citet{DeMeoDB} taxonomy -- are intermediate between S and A, and S and R, respectively. The visible (0.44--0.92\,$\mu$m) spectra that define these classes have a very steep ultraviolet slope shortward of 0.7\,$\mu$m. The 1\,$\mu$m absorption feature while deep and clearly visible in S$_{\rm r}$,
is shallow and not well defined in S$_{\rm a}$, which shows that it might be shifted above 1\,$\mu$m 
and can be associated with a presence of olivine. To compute colours representing the S$_{\rm a}$ 
and S$_{\rm r}$ classes we searched for asteroids belonging to those classes in the SMASS II data base. SDSS colours 
for five S$_{\rm a}$ and three S$_{\rm r}$ objects included in the SDSS data base of moving objects \citep[EAR-A-I0035-5-SDSSTAX-V1.1;][]{SaSr} 
were converted into the {\it VRI} system using the same procedure as for the ACAM measurements. 

Our colours for the Main Belt asteroids (289) Nenetta and (3819) Robinson are included as well. Both of these asteroids have 
been mineralogically classified as olivine-dominated [defined as S(I)-type in the \citet{STypes} classification] based on 
their 0.5--2.5 $\mu$m spectra \citep{Sanchez}. Based on visible (0.44--0.92 $\mu$m) spectra alone, \citet{BusBinzel} 
classified (289) Nenetta as an A-type asteroid and (3819) Robinson as an S$_{\rm r}$-type asteroid. Interestingly, Eureka was also classified as 
S$_{\rm r}$ in their work. Our $(V-R)$ colour of (3819) Robinson is lower than that of (289) Nenetta, consistent with the somewhat shallower 
slope shortward of the reflectance peak for that spectral class as compared to the A type.

We therefore conclude that the colour of (385250) 2001 DH47, after taking into account rotation-related photometric effects, 
is consistent with the spectroscopic determination of its taxonomy. Taking into account the asteroids' rotational brightness changes will be important in our future attempts to constrain the taxonomic class through colour photometry.

\section{Discussion and Conclusions}
\label{blabla}

Our spectroscopic and spectro-photometric data confirm that three members of the
Eureka family in the L5 Mars Trojan cloud -- including the previously-observed Eureka -- exhibit properties that are best 
interpreted in terms of a high surface abundance of olivine. Objects sharing the
same property, taxonomically classified as members of the A and S$_{\rm a}$ classes in the \citet{DeMeoDB} taxonomy, are 
quite unusual among the asteroid population. As mentioned in the Introduction, there are
reasons to believe that olivine-rich bodies were common among planetesimals accreted in 
the inner regions of the Solar System during the early phases of planetary formation.
The current underabundance of such objects may well indicate that most of those that were 
originally present have been lost -- so called "missing mantle problem" \citep{AIV}. 
If this interpretation is correct, then it is interesting that we find a small group 
of these bodies in one of the Mars Trojan clouds. Due to the particular dynamical 
environment that ensures orbital stability over long time-scales, the Martian Trojan clouds are 
one of the few places where one would expect to find samples of the first generation of planetesimals 
accreted in this region. In other words, these asteroids 
might well be samples of the original building blocks that came together to form Mars and of the other 
terrestrial planets. The common fate of these bodies elsewhere
seems to have been a nearly complete removal, possibly the result 
of intense collisional evolution \citep{Burbine.et.al1996,Chapman1997,Sanchez}. It is 
also possible that the dynamical excitation of planetesimals in the current main belt
during an early phase of migration of the giant planets \citep{Walsh.et.al2011, Walsh.et.al2012}
is implicated in the disappearance of these objects.

The fact that these olivine-rich asteroids belong to a group (the Eureka family) of
objects that may share a common origin is also interesting. Recent numerical modelling of the family's evolution under planetary gravitational perturbations and the Yarkovsky effect led \citet{Cuk.et.al2015} to conclude that the group is likely a genetic family formed roughly in the last Gyr of the Solar System's history. Whether or not the family sprung off a common parent -- a proto-Eureka -- has an obvious bearing on the relative
abundance of olivine-rich material near Mars in the early Solar System. It is also important to view this
in the context of the apparent compositional diversity of the Martian Trojan population overall.
\citet{Rivkin.et.al2003} obtained visible spectra of Eureka, (101429) 1998 VF31 at L5 and (121514) 1999 UJ7 at L4,
later complemented by NIR spectral coverage for the first two asteroids \citep{Rivkin.et.al2007}. They found that the latter two asteroids 
do not share Eureka's taxonomy and concluded that all these objects were once parts of larger bodies that formed separately in different locations of the Solar System. 

Future investigations of objects in Mars' Trojan clouds 
should include high-S/N spectra in the visual and NIR 
spectral regions to help us better understand the compositional relationships.
Extending the Mars Trojan inventory down to smaller sizes and determining their rotational 
characteristics would also help evaluate the stability of these objects in the size regime where non-gravitational perturbations
such as the Yarkovsky effect become important.  
 
%
%
%
%
%
\section*{Acknowledgements}

This work was supported via a grant (ST/M000834/1) from the UK Science and Technology Facilities Council.
Based on observations collected at the European Organisation for Astronomical Research in the 
Southern Hemisphere under ESO programme 296.C-5030 (PI: A. Christou). \\
We gratefully acknowledge observing grant support from the Institute of Astronomy and 
Rozhen National Astronomical Observatory, Bulgarian Academy of Sciences.\\
We thank Colin Snodgrass for kindly agreeing to observe asteroid 3819 with ACAM@WHT during program ITP6, 
and Maxime Devogele for his suggestion to compare our asteroid spectra with laboratory olivine spectra. \\
Astronomical research at the Armagh Observatory and Planetarium is grant-aided by the Northern Ireland Department for Communities (DfC). \\
This research utilises spectra acquired from the NASA textsc{relab} facility at Brown University.\\
Eureka spectral data utilised in this publication were obtained and made available by the The 
MIT-UH-IRTF Joint Campaign for NEO Reconnaissance. The IRTF is operated by the University 
of Hawaii under Cooperative Agreement no. NCC 5-538 with the National Aeronautics and Space 
Administration, Office of Space Science, Planetary Astronomy Program. The MIT component of 
this work is supported by NASA grant 09-NEOO009-0001, and by the National Science Foundation 
under Grants Nos. 0506716 and 0907766.




\bibliographystyle{mnras}
\bibliography{mars3} 








\bsp	
\label{lastpage}
\end{document}